\documentclass[12pt]{article}

\usepackage[T1]{fontenc}
\usepackage[utf8]{inputenc}

\usepackage{amsmath,amssymb,amsthm,mathtools}
\usepackage{bm}
\usepackage{subcaption}
\usepackage{graphicx}
\usepackage{array,booktabs,tabularx,multirow}
\usepackage{pdflscape}
\usepackage{enumitem}
\usepackage{appendix}
\usepackage{newunicodechar}
\newunicodechar{−}{-}
\usepackage{tikz}
\usetikzlibrary{matrix,positioning}
\usepackage{natbib}
\usepackage{setspace}
\usepackage{tocloft}
\usepackage[dvipsnames]{xcolor}

\usepackage[
    colorlinks = true,
    linkcolor  = blue, % section refs, equations, etc.
    citecolor  = blue,  % citations
    urlcolor   = blue      % URLs
]{hyperref}

\allowdisplaybreaks

\DeclareUnicodeCharacter{2234}{\therefore}
\DeclareUnicodeCharacter{202F}{~}

\newcolumntype{P}[1]{>{\raggedright\arraybackslash}p{#1}}
\newcolumntype{Y}{>{\raggedright\arraybackslash}X}

\usepackage[margin=1in]{geometry}

\usepackage{authblk}

\cftpagenumbersoff{figure}
\cftpagenumbersoff{table}

\newcommand{\keywords}[1]{%
  \vspace{0.5em}%
  \noindent\textbf{Keywords: }#1
}

\title{Recombination-Rate Modifiers Under Stochastic Transmission}
\author[1]{Elisa Heinrich-Mora}
\author[1]{Marcus Feldman\thanks{Corresponding author: mfeldman@stanford.edu}}
\affil[1]{Department of Biology, Stanford University, Stanford, CA, USA}

\begin{document}
% % -------- TITLE PAGE --------
\begin{titlepage}
    \thispagestyle{empty}
    \centering
     \vspace*{4em}

    {\LARGE Recombination Rate Modifiers under Stochastic Transmission\\}
    \vspace{1em}
    {\large Elisa Heinrich-Mora, Marcus Feldman\footnote{Corresponding author: mfeldman@stanford.edu}\par}
    \vspace{1em}
    {\small Department of Biology, Stanford University, Stanford, CA, USA\par}
    \vspace*{\fill}

\end{titlepage}

\begin{abstract}
The Reduction Principle states that, near a stable equilibrium under fixed viability selection, a selectively neutral modifier allele that reduces recombination rate among selected loci is favored, whereas one that increases recombination rate is eliminated. This result relies on the assumption that transmission parameters are constant across generations, so that invasion is governed by the dominant eigenvalue of a single transmission–selection matrix. Here we examine a minimal departure from this framework. In a three-locus diploid model, two loci experience symmetric multiplicative viability selection and a third, neutral locus modifies their recombination rate. All parameters are fixed except that recombination in modifier heterozygotes varies randomly across generations according to an i.i.d. stochastic process. When the recombination rate in modifier heterozygotes is constant, the Reduction Principle holds exactly: invasion occurs if the rare modifier allele reduces recombination relative to the resident rate. When recombination rate varies randomly across generations, invasion is governed by the top Lyapunov exponent of a product of random matrices. We show that temporal variation in recombination rate alone, in the absence of fluctuating viability selection, can reverse the direction of selection on the modifier locus predicted by the deterministic model. The mean recombination rate is insufficient to predict $M_2$ invasion; instead, outcomes depend on the full distribution of recombination rates and their ordered accumulation across generations. Parameters that affect only the magnitude of selection under constant transmission—including resident recombination, selection strength, and background linkage—can alter its sign under stochastic transmission. These results demonstrate that temporal variability in transmission constitutes an independent and qualitatively distinct force in the evolution of recombination rates.
\end{abstract}
\keywords{recombination rate modifiers; stochastic transmission; mutation--selection balance; reduction principle; recombination; evolutionary genetics}

\newpage
\doublespacing
\section*{Introduction}
Recombination shuffles genetic associations without altering allelic states themselves. It is therefore a force that may counteract selection by dismantling combinations it has already produced. Recombination has been at the center of population--genetic theory for more than a century \cite{fisher_genetical_1999,haldane_causes_1990,felsenstein_evolutionary_1974, svedOneHundredYears2018}. How recombination rate evolve is formalized in the theory of \emph{recombination rate modifiers}: loci whose alleles affect the rate of recombination among other loci, yet have no direct effect on fitness components at those loci \cite{feldman_population_1996}.

The standard modifier framework considers a population in which the genotype frequencies at a set of selected loci, called the major genes, are close to a locally stable equilibrium under fixed selection. One asks whether a rare allele at a modifier locus—whose sole effect is to change recombination rates among the major loci, and which has no direct effect on their fitnesses—can invade when introduced at low frequency. Under specific conditions on selection and linkage for the major loci, modifier alleles that reduce recombination rates increase in frequency, whereas those that increase recombination rates are eliminated \cite{feldman_selection_1972, karlin_towards_1974, felsenstein_evolutionary_1974, charlesworth_recombination_1976, feldman1976genetic, feldman_evolution_1980, feldman_evolutionary_1986, altenberg_unified_2017}. This is the \emph{Reduction Principle}. It is not a claim that low recombination rate is intrinsically advantageous, but a structural statement about dynamics near equilibrium. Selection generates linkage disequilibrium; recombination breaks it. A modifier allele that reduces this shuffling preserves the association structure on which selection is already acting, and is therefore indirectly favored.

The Reduction Principle rests on restrictive assumptions. When selection on the major genes varies through time, higher recombination rates may be favored \cite{charlesworth_selection_1979, otto_evolution_1997, lenormand_evolution_2000, carjaEvolutionStochasticFitnesses2013}. In such cases, selection no longer acts on a static system near equilibrium; instead, the population is repeatedly displaced from it, and recombination alters the subsequent dynamics. The evolutionary fate of recombination rate modifiers is then determined not by the preservation of established associations, but by how often transiently favorable combinations are generated.

But variability need not arise from selection. It may arise in the transmission process itself. Recently, we analyzed mutation-rate modifiers in which the mutation process varies across generations while viability selection remains fixed \cite{heinrich-moraEvolutionStochasticTransmission2026}. There, invasion was shown to be governed not by a single deterministic stability matrix but by the top Lyapunov exponent of a product of random matrices. Even small temporal variability in mutation could overturn deterministic results. That analysis established a stochastic modifier framework for transmission processes. 

Recombination rates are not fixed constants of genetic systems. Empirical studies show that they respond plastically to external conditions such as temperature, physiological stress, and seasonality, often synchronously across individuals within a generation \cite{plough_effect_1917, stern_effect_1926, modliszewski_elevated_2018}.  Such environmentally induced changes in recombination rates have been documented across different systems, and frequently occur without any detectable change in viability selection \cite{wilson_temperature_1959, stevison_recombination_2017}, which suggests that recombination rates can fluctuate at the population level even when the selective regime acting on phenotypes remains effectively constant.

Here we extend the stochastic modifier framework to recombination while holding viability selection fixed. We study a three-locus diploid model: two loci experience symmetric multiplicative viability selection; a third, neutral locus modifies the recombination rate between them. Under constant recombination, the Reduction Principle holds exactly. Invasion of a rare modifier allele is determined by the dominant eigenvalue of a fixed transmission–selection local stability matrix.

We then replace the constant recombination rate in modifier heterozygotes with stochastic variation. In each generation, all modifier heterozygotes in the population experience the same recombination rate, but this rate changes across generations according to a stochastic process, while viability selection remains fixed. In this setting there is no single matrix governing invasion. The modifier’s fate is determined by the long-run exponential growth rate of a product of generation-specific random matrices, that is, by the top Lyapunov exponent.

The result is precise. Temporal variation in recombination rate alone—without fluctuating selection—can alter the direction of selection on recombination rate modifiers. The mean recombination rate does not suffice; selection acts on the ordered accumulation of transmission events across generations. The Reduction Principle remains valid within its domain, but that domain excludes even minimal stochasticity in recombination rate. Once transmission parameters vary randomly through time, the deterministic conclusions no longer hold.

\section{Model Set-Up} \label{sec:model_setup}
We consider a large, randomly mating diploid population with discrete, non-overlapping generations. Three biallelic loci are arranged linearly along a chromosome in the order $\mathbf{A}$–$\mathbf{B}$–$\mathbf{M}$. Locus $\mathbf{A}$ has alleles $A_1,A_2$, locus $\mathbf{B}$ has alleles $B_1,B_2$, and the modifier locus $\mathbf{M}$ has alleles $M_1,M_2$. Viability selection acts only on loci $\mathbf{A}$ and $\mathbf{B}$. The modifier locus $\mathbf{M}$ is selectively neutral and affects the population solely through its influence on recombination rate. Therefore, the eight haplotypes are
\[
A_1B_1M_1,\; A_1B_2M_1,\; A_2B_1M_1,\; A_2B_2M_1,\;
A_1B_1M_2,\; A_1B_2M_2,\; A_2B_1M_2,\; A_2B_2M_2,
\]
with corresponding frequencies denoted by $x_1,\dots,x_8$. The gamete frequencies produced by each diploid genotype are shown in Appendix~\ref{appx:freqs_genotypes}.

\paragraph{Selection.} Viability depends only on the diploid genotypes at loci $\mathbf{A}$ and $\mathbf{B}$ and is identical for all modifier genotypes. We assume a symmetric multiplicative fitness scheme: genotypes homozygous at both selected loci have viability $(1-s)^2$, genotypes heterozygous at exactly one locus have viability $(1-s)$, and double heterozygotes have viability $1$, with $0<s<1$. This scheme treats the two selected loci symmetrically and assigns fitness according to the number of heterozygous loci.

Let $w_{ij}$ denote the relative fitness of the diploid genotype formed by haplotypes $i$ and $j$, determined solely by their alleles at $\mathbf{A}$ and $\mathbf{B}$. For example, $w_{11}$ is the fitness of genotype $A_1B_1/A_1B_1$, independent of the modifier alleles.

The marginal fitness of haplotype $i$ is
\[
    w_i = \sum_{j=1}^{8} w_{ij} x_j,
\]
the average fitness of haplotype $i$ when paired at random with the population. The mean fitness is
\[
    \bar{w} = \sum_{i=1}^{8} x_i w_i
            = \sum_{i=1}^{8} \sum_{j=1}^{8} w_{ij} x_i x_j.
\]
\paragraph{Recombination.} Recombination between loci $\mathbf{A}$ and $\mathbf{B}$ depends on the modifier genotype, occurring at rates $r_{11}$, $r_{12}$, and $r_{22}$ for genotypes $M_1M_1$, $M_1M_2$, and $M_2M_2$, respectively. Recombination between loci $\mathbf{B}$ and $\mathbf{M}$ occurs at a fixed rate $r$, independent of modifier genotype. The modifier therefore alters transmission of the selected loci without directly affecting viability.

\paragraph{Recursions.} Let $x_i$ and $x_i'$ denote the frequencies of haplotype $i$ in the current and next generation. Under the selection and recombination regime described above, the haplotype dynamics are given by Eqs.~\eqref{eq:1a}--\eqref{eq:1h} (see ref. \cite{feldman_selection_1972}):
\begin{align}
    \bar{w} x'_1 &= x_1 w_1 - r_{11} w_{14} (x_1 x_4 - x_2 x_3) - r_{12} (w_{17} x_1 x_7 + w_{18} x_1 x_8 - w_{35} x_3 x_5 - w_{36} x_3 x_6) \notag \\
    &\quad - r (w_{16} x_1 x_6 + w_{17} x_1 x_7 + w_{18} x_1 x_8 - w_{35} x_3 x_5 - w_{45} x_4 x_5 - w_{25} x_2 x_5) \notag \\
    &\quad + r r_{12} (2 w_{17} x_1 x_7 - 2 w_{35} x_3 x_5 + w_{18} x_1 x_8 - w_{45} x_4 x_5 + w_{27} x_2 x_7 - w_{36} x_3 x_6), \label{eq:1a} \\
    \bar{w} x'_2 &= x_2 w_2 - r_{11} w_{14} (x_2 x_3 - x_1 x_4) - r_{12}(w_{28} x_2 x_8 - w_{46} x_4 x_6 + w_{27} x_2 x_7 - w_{45} x_4 x_5) \notag \\
    &\quad - r (w_{25} x_2 x_5 - w_{16} x_1 x_6 + w_{27} x_2 x_7 + w_{28} x_2 x_8 - w_{36} x_3 x_6 - w_{46} x_4 x_6) \notag \\
    &\quad + r r_{12} (w_{28} x_1 x_8 + w_{27} x_2 x_7 + 2 w_{28} x_2 x_8 - 2 w_{46} x_4 x_6 - w_{36} x_3 x_6 - w_{45} x_4 x_5), \notag \\
    \bar{w} x'_3 &= x_3 w_3 - r_{11} w_{14} (x_2 x_3 - x_1 x_4) - r_{12}(-w_{17}x_1 x_7 - w_{18}x_1 x_8 + w_{35} x_3 x_5 + w_{36} x_3 x_6) \notag \\
    &\quad - r (w_{35} x_3 x_5 + w_{36} x_3 x_6 + w_{38} x_3 x_8 - w_{17} x_1 x_7 - w_{27} x_2 x_7 - w_{47} x_4 x_7) \notag \\
    &\quad + r r_{12} (2 w_{35} x_3 x_5 - 2 w_{17} x_1 x_7 - w_{18} x_1 x_8 - w_{27} x_2 x_7 + w_{36} x_3 x_6 + w_{45} x_4 x_5), \notag \\
    \bar{w} x'_4 &= x_4 w_4 - r_{11} w_{14} (x_1 x_4 - x_2 x_3) - r_{12}(w_{45} x_4 x_5 + w_{46} x_4 x_6 - w_{27} x_2 x_7 - w_{28} x_2 x_8) \notag \\
    &\quad - r (w_{45} x_4 x_5 + w_{46} x_4 x_6 + w_{47} x_4 x_7 - w_{18} x_1 x_8 - w_{28} x_2 x_8 - w_{38} x_3 x_8) \notag \\
    &\quad + r r_{12} (2 w_{46} x_4 x_6 - 2 w_{28} x_2 x_8 + w_{36} x_3 x_6 + w_{45} x_4 x_5 - w_{18} x_1 x_8 - w_{27} x_2 x_7), \notag \\
    \bar{w} x'_5 &= x_5 w_5 - r_{22} w_{58} (x_5 x_8 - x_6 x_7) - r_{12}(-w_{17} x_1 x_7 - w_{27} x_2 x_7 + w_{35} x_3 x_5 + w_{45} x_4 x_5) \notag \\
    &\quad - r (w_{25} x_2 x_5 - w_{16} x_1 x_6 - w_{17} x_1 x_7 - w_{18} x_1 x_8 + w_{35} x_3 x_5 + w_{45} x_4 x_5) \notag \\
    &\quad + r r_{12} (2 w_{35} x_3 x_5 - 2 w_{17} x_1 x_7 - w_{18} x_1 x_8 - w_{27} x_2 x_7 + w_{36} x_3 x_6 + w_{45} x_4 x_5), \notag \\
    \bar{w} x'_6 &= x_6 w_6 - r_{22} w_{58} (x_6 x_7 - x_5 x_8) - r_{12}(- w_{18} x_1 x_8 - w_{28} x_2 x_8 + w_{36} x_3 x_6 + w_{46} x_4 x_6) \notag \\
    &\quad - r ( w_{16} x_1 x_6 - w_{25} x_2 x_5 - w_{27} x_2 x_7 - w_{28} x_2 x_8 + w_{36} x_3 x_6 + w_{46} x_4 x_6) \notag \\
    &\quad + r r_{12} (2 w_{46} x_4 x_6 - 2 w_{28} x_2 x_8 - w_{18} x_1 x_8 - w_{27} x_2 x_7 + w_{36} x_3 x_6 + w_{45} x_4 x_5), \notag \\
    \bar{w} x'_7 &= x_7 w_7 - r_{22} w_{58} (x_6 x_7 - x_5 x_8) - r_{12}(w_{17} x_1 x_7 + w_{27} x_2 x_7 - w_{35} x_3 x_5 - w_{45} x_4 x_5) \notag \\
    &\quad - r (w_{17} x_1 x_7 + w_{27} x_2 x_7 - w_{35} x_3 x_5 - w_{36} x_3 x_6 + w_{47} x_4 x_7 - w_{38} x_3 x_8) \notag \\
    &\quad + r r_{12} (2 w_{17} x_1 x_7 - 2 w_{35} x_3 x_5 + w_{18} x_1 x_8 + w_{27} x_2 x_7 - w_{36} x_3 x_6 - w_{45} x_4 x_5), \notag \\
    \bar{w} x'_8 &= x_8 w_8 - r_{22} w_{58} (x_5 x_8 - x_6 x_7) - r_{12}(w_{18} x_1 x_8 + w_{28} x_2 x_8 - w_{36} x_3 x_6 - w_{46} x_4 x_6) \notag \\
    &\quad - r (w_{18} x_1 x_8 + w_{28} x_2 x_8 + w_{38} x_3 x_8 - w_{45} x_4 x_5 - w_{46} x_4 x_6 - w_{47} x_4 x_7) \notag \\
    &\quad + r r_{12} (w_{18} x_1 x_8 + w_{27} x_2 x_7 + 2 w_{28} x_2 x_8 - w_{36} x_3 x_6 - w_{45} x_4 x_5 - 2 w_{46} x_4 x_6). \label{eq:1h}
\end{align}

\medskip
\paragraph{Resident equilibrium.} We first find a stable equilibrium of the resident population when the modifier locus is fixed for $M_1$. In this case all individuals are $M_1M_1$, recombination between the selected loci $\mathbf{A}$ and $\mathbf{B}$ occurs at rate $r_{11}$, and the system reduces to the standard two-locus viability-selection model for haplotypes $A_1B_1$, $A_1B_2$, $A_2B_1$, and $A_2B_2$, with frequencies $x_1,\dots,x_4$.

Let $p=x_1+x_2$ and $q=x_1+x_3$ denote the allele frequencies of $A_1$ and $B_1$, and let $D=x_1x_4-x_2x_3$ denote linkage disequilibrium between the $A$ and $B$ loci; with $x_1=pq+D$, $x_2=p(1-q)-D$, $x_3=(1-p)q-D$, and $x_4=(1-p)(1-q)+D$.

Under symmetric multiplicative selection the two-locus dynamics are the same under interchange of the coupling haplotypes $A_1B_1 \leftrightarrow A_2B_2$ and the repulsion haplotypes $A_1B_2 \leftrightarrow A_2B_1$. The subspace $x_1=x_4$, $x_2=x_3$ is therefore invariant. It contains the internal polymorphic equilibria of the system and represents the equilibrium structure implied by fitness symmetry. Restricting attention to this manifold entails no loss of generality; there are no other equilibria.

With $x_1=x_4$, $x_2=x_3$, $p=q=\tfrac12$, so that $x_1=x_4=\tfrac14+D$ and $x_2=x_3=\tfrac14-D$, with $|D|\le\tfrac14$. Substitution into the two-locus recursion yields an equilibrium condition for $\hat D$ \cite{karlin_linkage_1970}, which reduces to
\[
\hat  D\Bigl(64\,s^2\hat  D^2-4(s^2-4r_{11})\Bigr)=0,
\]
whose solutions are $\hat D=0$ and $\hat D=\pm \tfrac14\sqrt{1-\tfrac{4r_{11}}{s^2}}$, with nonzero solutions existing if and only if $0<r_{11}<s^2/4$.

We focus on equilibria with $\hat D\neq0$, since recombination modifiers act only through their effect on existing genetic associations; at linkage equilibrium ($\hat D=0$) recombination has no first-order effect. The two equilibria with $\hat  D\neq0$ differ only in which allelic combinations are in excess. Without loss of generality, take
\[
\hat D=\tfrac14\sqrt{1-\tfrac{4r_{11}}{s^2}},
\qquad
0<r_{11}<\tfrac{s^2}{4},
\]
so that $\hat x_1=\hat x_4=\tfrac14+\hat D$ and $\hat x_2=\hat x_3=\tfrac14-\hat D$. In the full three-locus state space the resident equilibrium is
\[
\hat{\mathbf x}=(\hat x_1,\hat x_2,\hat x_3,\hat x_4,0,0,0,0),
\]
with mean fitness
\[
\hat{\bar w}=1-s+\tfrac{s^2}{4}+4s^2\hat D^2.
\]

\section{Invasion Analysis} \label{sec:invasion}
We analyze the fate of a rare modifier allele $M_2$ introduced into a population that lies in a neighborhood of the resident equilibrium $\hat{\mathbf x}$. The modifier locus is initially fixed for $M_1$, so recombination between $\mathbf A$ and $\mathbf B$ occurs at rate $r_{11}$. 

For $0<r_{11}<s^2/4$, the resident population is near one of the two internal equilibria with $\hat D\neq0$ specified above, and where
\[
\hat x_1=\hat x_4=\tfrac14+\hat D,\qquad
\hat x_2=\hat x_3=\tfrac14-\hat D,\qquad
\hat D=\pm \frac14\sqrt{1-\frac{4r_{11}}{s^2}},
\]
with resident mean fitness 
\[
\hat{\bar w}=1-s+\tfrac{s^2}{4}+4s^2\hat D^2.
\]
Because the equilibrium is locally stable, resident dynamics rapidly approach $\hat{\mathbf x}$. We therefore write
\[
x_i=\hat x_i+\epsilon_i, \qquad i=1,\dots,8,
\]
where $\epsilon_i$ are deviations satisfying $\sum_{i=1}^8 \epsilon_i=0$ and $|\epsilon_i|\ll 1$, reflecting that the state space is the frequency simplex.  

The invasion analysis proceeds by linearizing the full recursion in a neighborhood of $(\hat{\mathbf x},0)$—that is, near the equilibrium at which $M_2$ is absent—and retaining only terms that are first order in the modifier frequency and in the deviations $\epsilon_i$. Thus the initial fate of $M_2$ is determined by the local linearization of the transmission–selection system about the resident equilibrium.

\subsection{Constant recombination} ~\label{sec:constant-recomb}
Consider the case in which recombination in $M_1M_2$ heterozygotes is constant across generations \cite{charlesworth_recombination_1976, feldman_selection_1973, feldman_evolution_1980},  and equal to $r_{12}$. Introduce $M_2$ at very low frequency and let
\[
\mathbf x_{\mathrm{inv}}=(x_5,x_6,x_7,x_8)^\top
\]
denote the frequencies of haplotypes carrying $M_2$. Then $x_j=O(\varepsilon)$ for $j=5,\dots,8$, while $M_2M_2$ genotypes occur at order $O(\varepsilon^2)$ and may be neglected.

Linearizing \eqref{eq:1a}--\eqref{eq:1h} in a neighborhood of $(\hat{\mathbf x},0)$ yields
\[
\hat{\bar w}\,\mathbf x'_{\mathrm{inv}}
=
J(r_{12})\,\mathbf x_{\mathrm{inv}},
\]
where $J(r_{12})$ is the Jacobian of the invader subsystem evaluated at the resident equilibrium. Terms involving $r_{22}$ appear only in products of invader frequencies and thus enter at order $O(\varepsilon^2)$; consequently $J$ is independent of $r_{22}$. The resident recombination rate $r_{11}$ enters only through the equilibrium quantities $\hat D$ and $\hat{\bar w}$.

Invasion of the modifier allele is determined by the dominant eigenvalue $\lambda_{\max}(J/\hat{\bar w})$. The modifier allele $M_2$ increases when rare if 
\[
M_2 \text{ increases when rare } 
\Longleftrightarrow 
\rho\!\left(\frac{1}{\hat{\bar w}}J(r_{12})\right) > 1 .
\]
At the symmetric equilibrium,
\[
\hat w_1=\hat w_4=(1-\tfrac{s}{2})^2+s^2\hat D,
\qquad
\hat w_2=\hat w_3=(1-\tfrac{s}{2})^2-s^2\hat D,
\]
\[
\hat{\bar w}=1-s+\tfrac{s^2}{4}+4s^2\hat D^2,
\qquad
\hat D^2=\tfrac{1}{16}\!\left(1-\frac{4r_{11}}{s^2}\right).
\]
Evaluating \eqref{eq:1a}--\eqref{eq:1h} at $(\hat{\mathbf x},0,0,0,0)$ and collecting $O(\varepsilon)$ terms yields \eqref{eq:x5-final}--\eqref{eq:x8-final} (cf.\ \cite{feldman1976genetic}).
\begin{align}
\hat{\bar w}\,x'_5
&= \Bigl(
    \bigl[(1-s)^2 + 1\bigr]\hat x_1 + 2(1-s)\hat x_2
    + (r r_{12} - r - r_{12}) \hat x_1
    + (1-s)(2 r r_{12} - 2 r - r_{12}) \hat x_2
  \Bigr) x_5 \notag\\
&\quad
+ r\bigl((1-s)\hat x_1 + r_{12}\hat x_2\bigr) x_6 \notag\\
&\quad
+ \Bigl(
    (r + r_{12} - 2 r r_{12})(1-s)\hat x_1
    + r_{12}(1-r)\hat x_2
  \Bigr) x_7 \notag\\
&\quad
+ r(1-r_{12})\hat x_1\,x_8,
\label{eq:x5-final}
\\[0.5em]
\hat{\bar w}\,x'_6
&= r\bigl(r_{12}\hat x_1 + (1-s)\hat x_2\bigr) x_5 \notag\\
&\quad
+ \Bigl(
    2(1-s)\hat x_1 + \bigl[(1-s)^2 + 1\bigr]\hat x_2
    + (1-s)(2 r r_{12} - 2 r - r_{12}) \hat x_1
    + (r r_{12} - r - r_{12}) \hat x_2
  \Bigr) x_6 \notag\\
&\quad
+ r(1-r_{12})\hat x_2\,x_7 \notag\\
&\quad
+ \Bigl(
    r_{12}(1-r)\hat x_1
    + (r + r_{12} - 2 r r_{12})(1-s)\hat x_2
  \Bigr) x_8,
\label{eq:x6-final}
\\[0.5em]
\hat{\bar w}\,x'_7
&= \Bigl(
    r_{12}(1-r)\hat x_1
    + (r + r_{12} - 2 r r_{12})(1-s)\hat x_2
  \Bigr) x_5 \notag\\
&\quad
+ r(1-r_{12})\hat x_2\,x_6 \notag\\
&\quad
+ \Bigl(
    2(1-s)\hat x_1 + \bigl[(1-s)^2 + 1\bigr]\hat x_2
    + (1-s)(2 r r_{12} - 2 r - r_{12}) \hat x_1
    + (r r_{12} - r - r_{12}) \hat x_2
  \Bigr) x_7 \notag\\
&\quad
+ r\bigl(r_{12}\hat x_1 + (1-s)\hat x_2\bigr) x_8,
\label{eq:x7-final}
\\[0.5em]
\hat{\bar w}\,x'_8
&= r(1-r_{12})\hat x_1\,x_5 \notag\\
&\quad
+ \Bigl(
    (r + r_{12} - 2 r r_{12})(1-s)\hat x_1
    + r_{12}(1-r)\hat x_2
  \Bigr) x_6 \notag\\
&\quad
+ r\bigl((1-s)\hat x_1 + r_{12}\hat x_2\bigr) x_7 \notag\\
&\quad
+ \Bigl(
    \bigl[(1-s)^2 + 1\bigr]\hat x_1 + 2(1-s)\hat x_2
    + (r r_{12} - r - r_{12}) \hat x_1
    + (1-s)(2 r r_{12} - 2 r - r_{12}) \hat x_2
  \Bigr) x_8.
\label{eq:x8-final}
\end{align}
For compactness, define
\[
\begin{aligned}
\mathcal{A} &:= \bigl[(1-s)^2 + 1\bigr]\hat x_1
      + 2(1-s)\hat x_2
      + (r r_{12} - r - r_{12}) \hat x_1
      + (1-s)(2 r r_{12} - 2 r - r_{12}) \hat x_2,\\[0.2em]
\mathcal{B} &:= r\bigl((1-s)\hat x_1 + r_{12}\hat x_2\bigr),\\[0.2em]
\mathcal{C} &:= (r + r_{12} - 2 r r_{12})(1-s)\hat x_1
      + r_{12}(1-r)\hat x_2,\\[0.2em]
\mathcal{D} &:= r(1-r_{12})\hat x_1,\\[0.2em]
\mathcal{E} &:= r\bigl(r_{12}\hat x_1 + (1-s)\hat x_2\bigr),\\[0.2em]
\mathcal{F} &:= 2(1-s)\hat x_1
      + \bigl[(1-s)^2 + 1\bigr]\hat x_2
      + (1-s)(2 r r_{12} - 2 r - r_{12}) \hat x_1
      + (r r_{12} - r - r_{12}) \hat x_2,\\[0.2em]
\mathcal{G} &:= r(1-r_{12})\hat x_2,\\[0.2em]
\mathcal{H} &:= r_{12}(1-r)\hat x_1
      + (r + r_{12} - 2 r r_{12})(1-s)\hat x_2.
\end{aligned}
\]

From \eqref{eq:x5-final}–\eqref{eq:x8-final}, before rescaling by $\hat{\bar w}$, the Jacobian governing the linearized invasion dynamics is
\begin{equation}
J=
\begin{pmatrix}
\mathcal{A} & \mathcal{B} & \mathcal{C} & \mathcal{D}\\
\mathcal{E} & \mathcal{F} & \mathcal{G} & \mathcal{H}\\
\mathcal{H} & \mathcal{G} & \mathcal{F} & \mathcal{E}\\
\mathcal{D} & \mathcal{C} & \mathcal{B} & \mathcal{A}
\end{pmatrix}.
\label{eq:J-centro}
\end{equation}
The matrix $J$ is centrosymmetric, a direct consequence of the symmetry of the resident equilibrium. This symmetry does not require a symmetric introduction of $M_2$: any initial invader state decomposes uniquely into symmetric and antisymmetric components, which evolve independently. Define
\begin{equation} \label{change_var_Tmax}
    y_1=x_5+x_8,\quad
    y_2=x_6+x_7,\quad
    y_3=x_5-x_8,\quad
    y_4=x_6-x_7,
    \qquad
    \text{and }\mathbf y=T\,\mathbf x_{\mathrm{inv}},
\end{equation}
with
\begin{equation}
T=
\begin{pmatrix}
1 & 0 & 0 & 1\\
0 & 1 & 1 & 0\\
1 & 0 & 0 & -1\\
0 & 1 & -1 & 0
\end{pmatrix},
\qquad
T^{-1}=\frac12
\begin{pmatrix}
1 & 0 & 1 & 0\\
0 & 1 & 0 & 1\\
0 & 1 & 0 & -1\\
1 & 0 & -1 & 0
\end{pmatrix}.
\label{eq:Tandinv}
\end{equation}

In these coordinates the linearized system becomes
\[
\hat{\bar w}\,\mathbf y'=\widetilde J\,\mathbf y,
\qquad
\widetilde J=TJT^{-1}=
\begin{pmatrix}
J^+ & 0\\
0   & J^-
\end{pmatrix},
\]
where
\begin{equation}
J^+=
\begin{pmatrix}
\mathcal A+\mathcal D & \mathcal B+\mathcal C\\
\mathcal E+\mathcal H & \mathcal F+\mathcal G
\end{pmatrix},
\qquad
J^-=
\begin{pmatrix}
\mathcal A-\mathcal D & \mathcal B-\mathcal C\\
\mathcal E-\mathcal H & \mathcal F-\mathcal G
\end{pmatrix}.
\label{eq:JplusJminus}
\end{equation}
A complete derivation is given in \eqref{eq:JplusJminus} (Appendix~\ref{app:blockdiag}). Thus the asymptotic growth rate of the modifier allele is $\max\{\rho(J^+/\hat{\bar w}),\rho(J^-/\hat{\bar w})\}$, where $\rho(\cdot)$ denotes the spectral radius, i.e.\ the maximum modulus of the eigenvalues of the corresponding matrix. For the resident equilibrium considered here, the dominant eigenvalue lies in the symmetric block, so the invasion threshold is obtained from $J^+/\hat{\bar w}$. 

Writing $J^+=\bigl(\begin{smallmatrix}a^+&b^+\\ c^+&d^+\end{smallmatrix}\bigr)$, the boundary $\rho(J^+/\hat{\bar w})=1$ is equivalent to $\det(J^+-\hat{\bar w}I_2)=0$, i.e.
\[
\hat{\bar w}^2-\tau\,\hat{\bar w}+\Delta=0,
\qquad
\tau=a^++d^+,\ \ \Delta=a^+d^+-b^+c^+.
\]
Substituting $\hat x_1=\tfrac14+\hat D$ and $\hat x_2=\tfrac14-\hat D$ into $J^+$ yields $a^+=a_0+a_1\hat D$, $d^+=a_0-a_1\hat D$, $b^+=b_0+b_1\hat D$, and $c^+=b_0-b_1\hat D$, with coefficients depending only on $(s,r,r_{12})$. Simplification gives
\[
\hat{\bar w}^2-\tau\,\hat{\bar w}+\Delta
=\hat D^2 s^2\bigl(16\hat D^2 s^2+4r_{12}-s^2\bigr).
\]
When $\hat D=0$ this expression vanishes identically and the linear analysis is inconclusive. Since $s>0$ and $\hat D^2>0$, invasion occurs if $16\hat D^2 s^2+4r_{12}-s^2<0$, equivalently $r_{12}<\tfrac{s^2}{4}(1-16\hat D^2)$. Using $\hat D^2=\tfrac{1}{16}\bigl(1-\tfrac{4r_{11}}{s^2}\bigr)$ yields
\[
r_{12}<r_{11},
\qquad
0<r_{11}<\frac{s^2}{4}.
\]
Thus, under constant recombination and near this stable equilibrium, a rare modifier allele $M_2$ increases in frequency if it reduces the recombination rate relative to the resident population. Modifiers that increase recombination are eliminated. This is the \emph{Reduction Principle} \cite{feldman_selection_1972, feldman_evolution_1980, karlin_towards_1974, charlesworth_recombination_1976}: holding selection parameters fixed, natural selection favors genetic modifier alleles that reduce the rate at which recombination disrupts the associations already generated by selection. The criterion depends on the resident linkage disequilibrium through $\hat D^2$ and is therefore independent of its sign and of $r$.

\subsection{Invasion by Stochastic recombination} 
Now suppose that recombination in $M_1M_2$ heterozygotes varies randomly across generations. In generation $t$ recombination occurs with probability $r_{12,t}$, where $\{r_{12,t}\}_{t\ge0}$ is an i.i.d.\ sequence of random variables supported on $[0,\tfrac12]$. All other aspects of the life cycle—including viability selection $s$, background recombination $r$, and recombination in the resident population $r_{11}$—are time invariant. It is assumed that all $M_1M_2$ individuals in a given generation $t$ produce $r_{12,t}$.

Because recombination among resident $M_1M_1$ individuals occurs at the fixed rate $r_{11}$, the resident equilibrium $\hat{\mathbf x}$ and its associated linkage disequilibrium
\[
\hat D=\pm \frac14\sqrt{1-\frac{4r_{11}}{s^2}}
\]
are unchanged from the constant-$r_{12}$ case. Temporal variation enters only through the modifier heterozygotes and does not perturb the resident state.

Linearizing the full recursions about $(\hat{\mathbf x},0,0,0,0)$ again yields a linear invasion system, now with random coefficients:
\begin{equation}
\mathbf x_{\mathrm{inv},t+1}
=\frac{1}{\hat{\bar w}}\,J(r_{12,t})\,\mathbf x_{\mathrm{inv},t},
\qquad
\mathbf x_{\mathrm{inv},t}=(x_{5,t},x_{6,t},x_{7,t},x_{8,t})^\top,
\label{eq:stoch-lin-inv}
\end{equation}
where $J(r_{12,t})$ is obtained from the deterministic Jacobian \eqref{eq:J-centro} by replacing $r_{12}$ with $r_{12,t}$. As before, terms involving $r_{22}$ enter only at quadratic order in invader frequencies and are therefore irrelevant for local dynamics.

For every realization of $r_{12,t}$, the matrix $J(r_{12,t})$ retains the same centrosymmetric structure as in the constant case \eqref{eq:J-centro}. This reflects the symmetry of the resident equilibrium and does not rely on any symmetry in the stochastic process itself. Consequently, the linear change of variables in \eqref{change_var_Tmax} block-diagonalizes $J(r_{12,t})$ for all $t$:
\[
T\,J(r_{12,t})\,T^{-1}
=
\begin{pmatrix}
J^+(r_{12,t}) & 0\\
0 & J^-(r_{12,t})
\end{pmatrix}.
\]
Writing $\mathbf y_t=T\mathbf x_{\mathrm{inv},t}$, the invasion dynamics become
\begin{equation}
\mathbf y_{t+1}
=\frac{1}{\hat{\bar w}}
\begin{pmatrix}
J^+(r_{12,t}) & 0\\
0 & J^-(r_{12,t})
\end{pmatrix}
\mathbf y_t.
\label{eq:stoch-block}
\end{equation}
Equation \eqref{eq:stoch-block} shows that the symmetric subspace, spanned by $(y_1,y_2)$, and the antisymmetric subspace, spanned by $(y_3,y_4)$, are invariant at every generation. This invariance holds almost surely and does not depend on the realization of $\{r_{12,t}\}$.

If the modifier allele $M_2$ is introduced symmetrically — $x_{5,0}=x_{8,0}$ and $x_{6,0}=x_{7,0}$, equivalently $y_{3,0}=y_{4,0}=0$ — then $(y_{3,t},y_{4,t})\equiv(0,0)$ for all $t$ with probability one. As in the constant case, this restriction entails no loss of generality: antisymmetric components correspond to differences between symmetry-related haplotypes and cannot dominate invasion when the resident equilibrium itself is symmetric. Invasion is therefore governed entirely by the symmetric subsystem
\begin{equation}
\mathbf y_{t+1}
=\frac{1}{\hat{\bar w}}\,J^+(r_{12,t})\,\mathbf y_t,
\qquad
\mathbf y_t=(y_{1,t},y_{2,t})^\top,
\label{eq:stoch-symm}
\end{equation}
where
\begin{equation}
J^+(r_{12,t})=
\begin{pmatrix}
a_0(r_{12,t})+a_1(r_{12,t})\hat D &
b_0(r_{12,t})+b_1(r_{12,t})\hat D\\
b_0(r_{12,t})-b_1(r_{12,t})\hat D &
a_0(r_{12,t})-a_1(r_{12,t})\hat D
\end{pmatrix},
\label{eq:Jplus-stoch}
\end{equation}
and the coefficients
\begin{align}
a_1(r_{12,t}) &= 2r\,(r_{12,t}-1)(s-1) + s\,(s-r_{12,t}), \notag\\
b_1(r_{12,t}) &= 2r\,(r_{12,t}-1)(s-1) - s\,r_{12,t}, \notag\\
a_0(r_{12,t}) &= -\tfrac14 a_1(r_{12,t})
               + \tfrac12\bigl((s-1)^2+1-r_{12,t}\bigr), \notag\\
b_0(r_{12,t}) &= \tfrac14\bigl(b_1(r_{12,t}) + 2 r_{12,t}\bigr)
\label{eq:ab-coefs}
\end{align}
are linear functions of $r_{12,t}$ and otherwise depend only on the fixed parameters $s$ and $r$.

The long-run fate of the modifier is determined by the asymptotic growth rate of the random matrix product associated with \eqref{eq:stoch-symm}. By the Furstenberg--Kesten theorem, the limit
\begin{equation}
\gamma=\lim_{t\to\infty}\frac{1}{t}
\log\Bigl\|
\tfrac{1}{\hat{\bar w}}J^+(r_{12,t-1})\cdots
\tfrac{1}{\hat{\bar w}}J^+(r_{12,0})\,\mathbf y_0
\Bigr\|
\label{eq:lyapunov_main}
\end{equation}
exists almost surely for any initial vector $\mathbf y_0>0$ and is independent of $\mathbf y_0$. The modifier allele $M_2$ invades if $\gamma>0$.

\medskip
\subsection{Comparison: constant vs stochastic recombination}
The difference between constant and temporally varying recombination rates is not merely quantitative but structural. When the recombination rate in $M_1M_2$ heterozygotes is constant, invasion of a rare modifier allele is governed by a single linear operator $J^{+}(r_{12})$ evaluated at the resident equilibrium. Writing $\lambda(r_{12})$ for its dominant eigenvalue and $\hat{\bar w}$ for resident mean fitness, invasion occurs if $\lambda(r_{12})/\hat{\bar w}>1$, which reduces exactly to $r_{12}<r_{11}$. Modifier alleles that increase recombination rate are eliminated, alleles that leave recombination unchanged are neutral, and—conditional on existence of the internal polymorphic equilibrium—all other parameters affect only the magnitude of the leading eigenvalue, not whether it is greater or less than $1$.

\begin{figure}[t]
  \centering
  \includegraphics[width=0.78\textwidth]{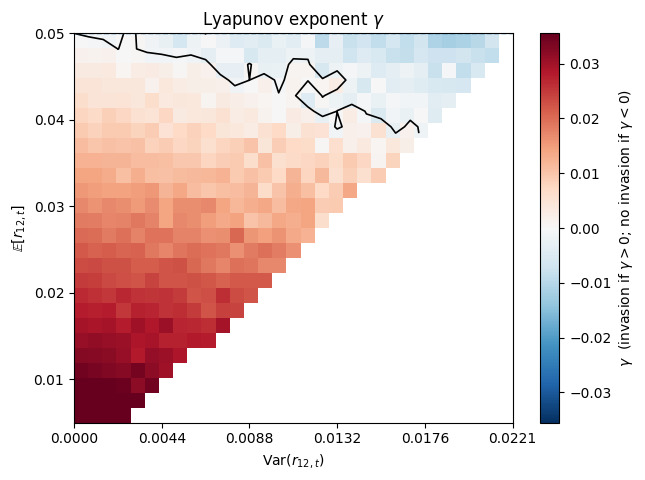}
 \caption{Top Lyapunov exponent $\gamma$ governing invasion of a rare recombination modifier allele under temporally varying recombination in $M_1M_2$ heterozygotes. In each generation, $r_{12,t}$ is drawn i.i.d.\ from a scaled Beta distribution on $[0,\tfrac12]$. The vertical axis shows the mean $\mathbb{E}[r_{12,t}]$ (restricted to $\mathbb{E}[r_{12,t}]\le r_{11}$), and the horizontal axis shows the variance $\mathrm{Var}(r_{12,t})$; blank cells indicate infeasible mean--variance combinations for this distribution. Colors denote the magnitude and sign of $\gamma$, with deeper orange indicating larger positive values (stronger invasion), deeper blue indicating more negative values (stronger elimination), and lighter shades indicating weaker effects. The black line marks $\gamma=0$. Parameters are $s=0.8$, $r=0.12$, and $r_{11}=0.05$, for which the resident polymorphism is locally stable with $\hat D\neq0$ and $\hat{\bar w}=0.47$.}
\label{fig:heatmap_mean_var}
\end{figure}

When recombination in $M_1M_2$ heterozygotes varies randomly across generations, $r_{12,t}$, the invasion dynamics are fundamentally altered. Growth of a rare modifier allele is governed by a product of random matrices $J^{+}(r_{12,t})$, and the relevant quantity is the top Lyapunov exponent $\gamma$ of this product. Invasion occurs if $\gamma>0$, and the sign and magnitude of $\gamma$ depend on the full distribution of recombination rates across generations, which cannot, in general, be inferred from their mean alone. Figure~\ref{fig:heatmap_mean_var} illustrates this dependence: deeper orange regions correspond to larger positive values of $\gamma$ (stronger invasion), deeper blue regions to more negative values (stronger elimination), and the black line marks $\gamma=0$. Along the vertical axis, the mean recombination rate of $M_1M_2$ is held below the resident rate $r_{11}=0.05$, for which the constant model predicts invasion throughout. However, as temporal variance increases, the Lyapunov exponent crosses zero and invasion fails. Temporal variability in transmission can therefore reverse the prediction of the constant model.

\medskip
With constant transmission, selection compares fixed recombination rates. With temporally varying transmission, selection acts on their ordered accumulation across generations. This distinction underlies all subsequent parameter effects.

\begin{figure}[t]
    \centering
    \includegraphics[width=1.1\textwidth]{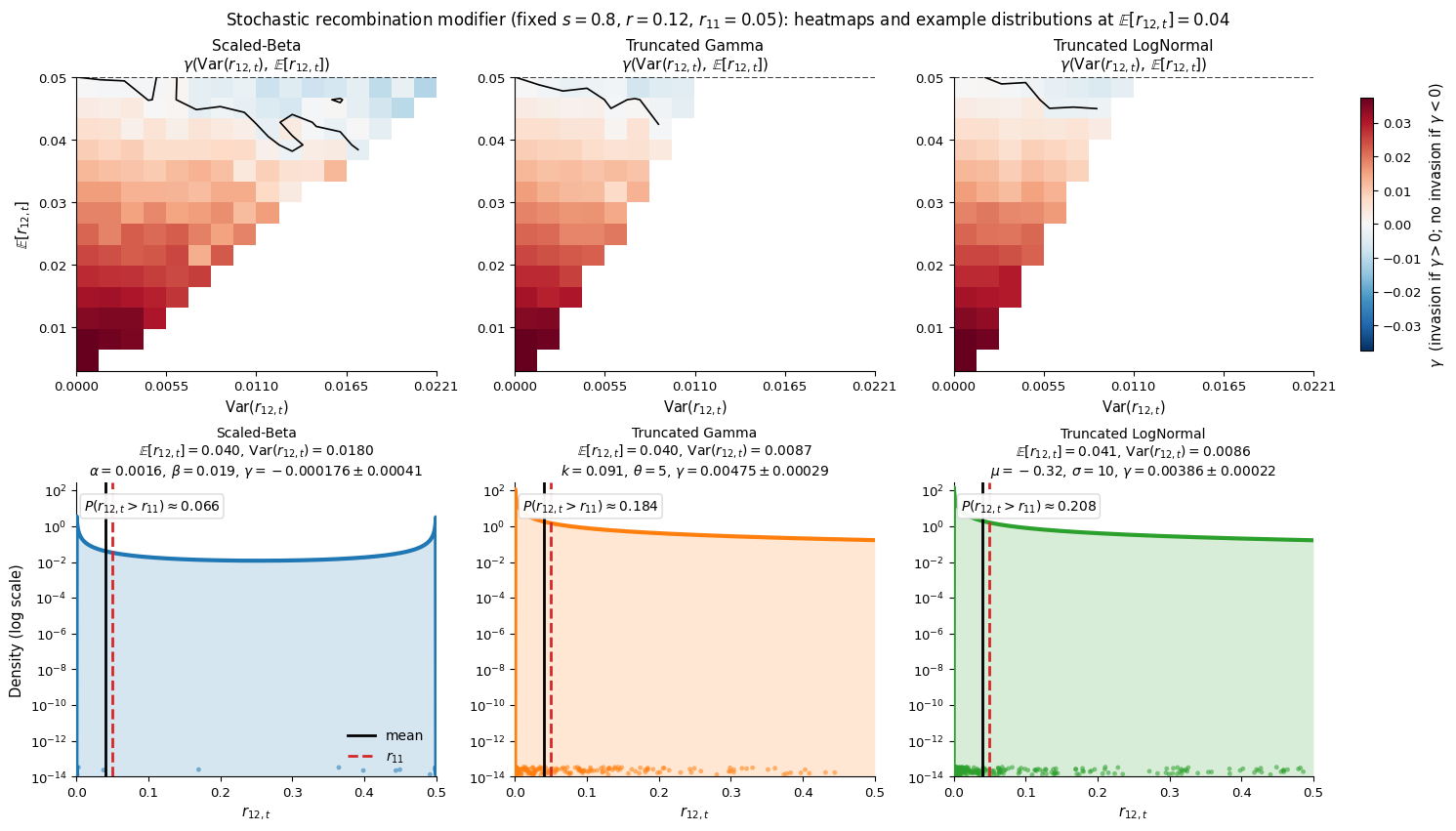}
    \caption{\textbf{Distributional effects on stochastic invasion of a recombination modifier.}
    \emph{Top row:} Heatmaps of the Lyapunov exponent $\gamma$ as a function of $\mathbb{E}[r_{12,t}]$ and $\mathrm{Var}(r_{12,t})$ for three distributions supported on $[0,\tfrac12]$ (scaled Beta, truncated Gamma, truncated LogNormal; truncated distributions are moment-matched in the truncated mean and variance). Colors indicate invasion – more orange ($\gamma>0$) – or extinction – more blue ($\gamma<0$); the black line marks $\gamma=0$. Blank cells correspond to infeasible mean--variance pairs. Other parameters are fixed at $s=0.8$, $r_{11}=0.05$, and $r=0.12$. \emph{Bottom row:} Example truncated distributions with matched mean $\mathbb{E}[r_{12,t}]=0.040<r_{11}$ and (approximately) maximal feasible variance within each distribution. Shaded regions indicate probability mass. Vertical lines mark $\mathbb{E}[r_{12,t}]$ (black) and $r_{11}$ (red dashed). The top-left labels reports the tail probability $P(r_{12,t}>r_{11})$ estimated from i.i.d.\ samples.}
    \label{fig:stochastic_tail_comparison}
\end{figure}

\begin{itemize}
    \item \emph{Heterozygote recombination $(r_{12}$ vs. $\{r_{12,t}\})$.} In the constant model, recombination in $M_1M_2$ heterozygotes is summarized by a single scalar $r_{12}$, and the invasion condition depends only on the sign of $r_{12}-r_{11}$. In the stochastic model, recombination is described by the entire distribution of $\{r_{12,t}\}$, and invasion depends on the top Lyapunov exponent of the associated random matrix product. Consequently, the mean $\mathbb E[r_{12,t}]$ is generally insufficient to determine the outcome.
    
    Figure~\ref{fig:stochastic_tail_comparison} shows how the temporal distribution of recombination rates affects invasion under stochastic transmission. The top row plots the Lyapunov exponent $\gamma$ over the $(\mathbb E[r_{12,t}],\mathrm{Var}(r_{12,t}])$ plane for three distribution families supported on $[0,\tfrac12]$. Blank cells indicate infeasible mean–variance combinations, either because they violate the variance bound for distributions on the interval or because the truncated family cannot attain them (within numerical tolerance). Throughout, $\mathbb E[r_{12,t}]<r_{11}$, a regime in which the constant-recombination model would predict invasion ($\gamma>0$). Increasing variance alone can reverse the sign of $\gamma$: sufficiently strong temporal fluctuations suppress invasion even when the mean recombination rate favors it.
  
    The bottom row of Figure \ref{fig:stochastic_tail_comparison} highlights the mechanism at fixed mean $\mathbb E[r_{12,t}]=0.040<r_{11}$, choosing (approximately) the maximal feasible variance within each family. Although the means coincide, the distributions differ in how probability mass is allocated relative to the resident rate $r_{11}$. Invasion is favored when most mass lies below $r_{11}$, so that generations with reduced recombination dominate the multiplicative growth process. It is suppressed when enough mass lies above $r_{11}$, because occasional high-recombination generations impose multiplicative losses that compound through time. The scaled Beta distribution places substantial mass near both boundaries and generates the largest upper-tail probability $P(r_{12,t}>r_{11})$, producing the strongest reduction in $\gamma$. The truncated Gamma and LogNormal distributions concentrate more mass near zero and exhibit lighter upper tails; their impact on $\gamma$ is correspondingly weaker, though still sufficient to prevent invasion when the variance is large enough.

\begin{figure}[t]
  \centering
  \includegraphics[width=0.85\textwidth]{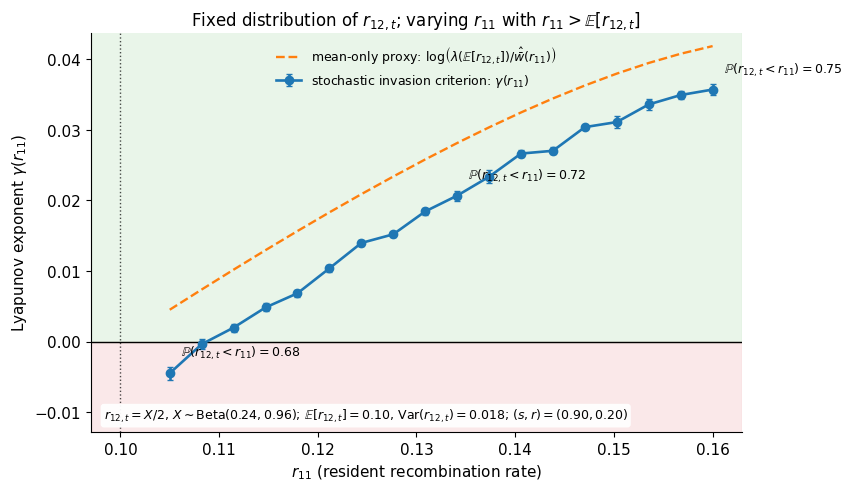}
\caption{\textbf{$r_{11}$ changes both the benchmark and the matrices.}
  Lyapunov exponent $\gamma(r_{11})$ (solid) for invasion under a fixed distribution of heterozygote recombination rates $\{r_{12,t}\}$; invasion occurs if $\gamma(r_{11})>0$. Here $r_{12,t}=X_t/2$ with $X_t\stackrel{\mathrm{i.i.d.}}{\sim}\mathrm{Beta}(a,b)$ on $[0,1]$, so the distribution of $r_{12,t}\in[0,\tfrac12]$ is held fixed as $r_{11}$ varies. The dashed curve is a mean-only proxy $\log(\lambda(\mathbb{E}[r_{12,t}])/\hat{\bar w}(r_{11}))$. Annotations at different points report $\mathbb{P}(r_{12,t}<r_{11})$, that is, the fraction of generations in which the recombination rate falls below the resident rate. Parameters are $(s,r)=(0.90,0.20)$, and only values with $0<r_{11}<s^{2}/4$ are shown.}
  \label{fig:r11_stochastic_fixed_dist}
\end{figure}
    \item \emph{Resident recombination rate $(r_{11})$.} Figure~\ref{fig:r11_stochastic_fixed_dist} isolates the effect of the resident recombination rate $r_{11}$ in the stochastic model by holding the distribution of $\{r_{12,t}\}$ fixed while varying $r_{11}$. In the constant model, $r_{11}$ enters only as the reference against which a constant $r_{12}$ is compared: the invasion boundary is always $r_{12}=r_{11}$, so changing $r_{11}$ cannot change the \emph{direction} of selection for a given ordering $r_{12}\lessgtr r_{11}$. It changes the resident equilibrium (through $\hat D$ and $\hat{\bar w}$) and therefore the \emph{rate} of invasion or loss, but not its sign.

    In the stochastic model, $r_{11}$ plays two logically distinct roles. First, it sets the moving threshold that partitions generations into those with $r_{12,t}<r_{11}$ and those with $r_{12,t}>r_{11}$. Increasing $r_{11}$ therefore increases $\mathbb P(r_{12,t}<r_{11})$ under a fixed distribution and shifts the balance of favorable versus unfavorable generations (as annotated in the figure). Second, $r_{11}$ determines the resident state itself: $\hat D=\hat D(r_{11})$ and $\hat{\bar w}=\hat{\bar w}(r_{11})$ enter every one-generation factor $\hat{\bar w}(r_{11})^{-1}J^{+}(r_{12,t};\hat D(r_{11}))$. Thus, changing $r_{11}$ not only reclassifies realizations of $r_{12,t}$; it changes the matrices being multiplied. 

    Figure~\ref{fig:r11_stochastic_fixed_dist}  shows the consequence of these two effects acting together. Although $r_{11}>\mathbb E[r_{12,t}]$ throughout, the Lyapunov exponent crosses from negative to positive as $r_{11}$ increases: a modifier allele whose mean effect is to reduce recombination can fail to invade when too much probability mass lies above the resident reference level, and can invade once the balance shifts. The dashed mean-only proxy, which ignores temporal ordering and treats transmission as constant at $\mathbb E[r_{12,t}]$, does not capture this transition. In this regime, $r_{11}$ is not merely a benchmark rate; it is a parameter that reshapes both the composition of favorable versus unfavorable generations and the strength with which each generation contributes to long-run growth.
  
\begin{figure}[t]
  \centering
  \includegraphics[width=0.85\textwidth]{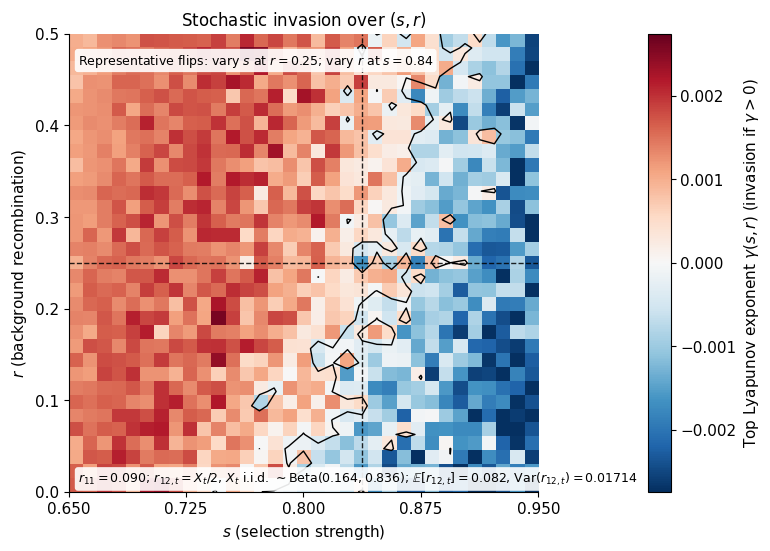}
  \caption{
  \textbf{Dependence of stochastic model on $s$ and $r$.} Heatmap of the top Lyapunov exponent $\gamma(s,r)$ for the linear recursion $\mathbf y_{t+1}=\hat{\bar w}(s,r,r_{11})^{-1}J^{+}(r_{12,t};s,r,r_{11})\,\mathbf y_t$, with $r_{11}=0.09$ and $r_{12,t}=X_t/2$, where $(X_t)$ are i.i.d.\ $\mathrm{Beta}(0.164,0.836)$ on $[0,1]$. Colors indicate the sign and magnitude of $\gamma$ (invasion if $\gamma>0$); the black line marks $\gamma=0$. Only parameter values satisfying $0<r_{11}<s^{2}/4$ are shown.
  }
  \label{fig:lyapunov_heatmap_sr}
\end{figure}

  \item \emph{Selection strength $(s)$ and background recombination parameter $(r)$.} The selection coefficient $s$ and the background recombination parameter $r$ enter the invasion problem in both the constant and stochastic models through the resident equilibrium and the transmission--selection matrix. In the constant model, conditional on existence of the polymorphic equilibrium, varying $s$ or $r$ changes the magnitude of the dominant eigenvalue but not whether it is greater or less than unity: the invasion boundary remains fixed at $r_{12}=r_{11}$, so the direction of selection on $M_2$ is unaffected once $r_{12}\lessgtr r_{11}$ is specified.
  
  In the stochastic model, $s$ and $r$ affect invasion in a fundamentally different way. Both parameters enter every realized one-generation matrix $J^{+}(r_{12,t};s,r,r_{11})$ and therefore influence the entire random product that determines the Lyapunov exponent. When realizations of $r_{12,t}$ fall on both sides of $r_{11}$ with positive probability, changing $s$ or $r$ alters not only the strength but also the relative contribution of favorable and unfavorable generations, and can reverse the sign of $\gamma$. Figure~\ref{fig:lyapunov_heatmap_sr} illustrates this dependence. The heat map shows the Lyapunov exponent $\gamma(s,r)$ for fixed $r_{11}$ and a fixed distribution of $\{r_{12,t}\}$, with colors indicating invasion ($\gamma>0$) or loss ($\gamma<0$). The black line marks $\gamma=0$. Vertical and horizontal dashed lines highlight representative cross-sections: varying $s$ with fixed $r$ or varying $r$ with fixed $s$ can each move the system across the invasion boundary, even though the distribution of heterozygote recombination is unchanged. Thus, parameters that are neutral with respect to the invasion in the constant model become decisive under temporally varying transmission.

\end{itemize}

\medskip
\medskip
The roles of the parameters in the two models are summarized in Table~\ref{tab:comparison}. Under constant transmission, selection on recombination modifiers reduces to a local comparison of rates; the sign of invasion is determined by the deterministic Jacobian, yielding the Reduction Principle. Under temporally varying transmission, selection operates on products of random matrices, so growth depends on the entire sequence of recombination realizations rather than on any instantaneous value. Consequently, arbitrarily small temporal variance in recombination can overturn the prediction of the constant model and reverse the direction of selection, even when viability selection and the mean recombination rate are held fixed.

\begin{table}[t]
\centering
\caption{Role of parameters in $M_2$ invasion under constant and stochastic recombination rates}
\label{tab:comparison}
\begin{tabular}{p{0.27\textwidth} p{0.33\textwidth} p{0.33\textwidth}}
\toprule
 & \textbf{Constant recombination} & \textbf{Stochastic recombination} \\
\midrule
Transmission in $M_1M_2$ &
Fixed rate $r_{12}$ &
Random sequence $\{r_{12,t}\}$ (distribution on $[0,\tfrac12]$) \\
Governing object &
Single matrix $J^{+}(r_{12})$ &
Random product $\prod_{t} J^{+}(r_{12,t})$ \\
Invasion criterion &
$\lambda(r_{12})/\hat{\bar w}>1$ &
Top Lyapunov exponent $\gamma>0$ \\
Sufficient summary of $r_{12,t}$ &
$r_{12}$ alone &
Not determined by $\mathbb{E}[r_{12,t}]$; depends on distributional mass relative to $r_{11}$ and tail weight\\
Effect of $\mathrm{Var}(r_{12,t})$ &
(No temporal variation)&
Can change sign of invasion (Figs.~\ref{fig:heatmap_mean_var}, \ref{fig:stochastic_tail_comparison}) \\
Role of $r_{11}$ &
Sets boundary $r_{12}=r_{11}$ &
Sets threshold and changes matrices via $\hat D(r_{11}),\hat{\bar w}(r_{11})$ (Fig.~\ref{fig:r11_stochastic_fixed_dist}) \\
Role of $s$ and $r$ &
Affect magnitude only &
Can reverse sign by altering each factor in the product (Fig.~\ref{fig:lyapunov_heatmap_sr}) \\
Reduction Principle &
Holds exactly: invade if $r_{12}<r_{11}$&
Not generically valid: temporal variation can reverse outcome \\
\bottomrule
\end{tabular}
\end{table}

\section*{Discussion}
The standard theory of recombination modifiers entails that: near a stable equilibrium at the selected loci, with viability selection fixed, a selectively neutral modifier allele that changes only recombination rates among those loci is favored if it \emph{reduces} recombination rate and is disfavored if it \emph{increases} recombination rate. In this standard formulation, the \emph{Reduction Principle} states that, under specific conditions on selection, linkage, and mating system, the leading eigenvalue governing invasion is $>1$, if $r_{12}<r_{11}$, and $<1$ if $r_{12}>r_{11}$. This result relies on the deterministic specification: transmission is represented by a fixed matrix, so selection on recombination reduces to a pointwise comparison of constant rates. Conditional on existence of the internal equilibrium, all other parameters —selection strength, background linkage, resident recombination— affect only the \emph{magnitude} of the growth rate, not its sign.

Our analysis identifies an empirically motivated violation of this determinism—temporal variation in recombination rates experienced by modifier heterozygotes—and shows that the Reduction Principle does not generically extend to this setting. The resident population is unchanged: recombination rate among resident genotypes occurs at a fixed rate, so the resident equilibrium and its linkage disequilibrium are exactly those of the deterministic model. What changes is the nature of the invasion problem itself. Growth of a rare modifier allele is no longer governed by the dominant eigenvalue of a single matrix, but by the top Lyapunov exponent of a product of generation-specific transmission--selection matrices. The sign of selection on the modifier allele $M_2$ is therefore a property of a multiplicative process, not of a single-generation expectation. In this regime, the mean recombination rate is not a sufficient statistic.

In the constant model, conditional on equilibrium existence, the inequality $r_{12}<r_{11}$ is sufficient for invasion uniformly across admissible parameter values. In the stochastic model, the same inequality applied to $\mathbb E[r_{12,t}]$ does not determine the outcome. When recombination rate in heterozygotes sometimes falls below and sometimes exceeds the resident rate, the invader allele experiences both favorable and unfavorable generations. Because the relevant matrices do not commute, the sequence and variability of transmission events matter: rare but extreme high-recombination generations can dominate long-run growth even when they occur infrequently. Temporal variance in transmission alone —without any change in viability— can reverse the direction of selection.

Crucially, temporal variation also affects the \emph{roles of other parameters} in the invasion process. In the constant model, once the ordering $r_{12}\lessgtr r_{11}$ is fixed, parameters such as selection strength $s$, background recombination between the modifier and the selected loci $r$, and the resident recombination rate $r_{11}$ affect only the speed of invasion or loss. They cannot change its direction. Under stochastic transmission, these same parameters play a qualitative role. The resident recombination rate $r_{11}$ no longer serves merely as a fixed benchmark: it determines both the partition of generations into favorable and unfavorable regimes and the structure of each one-generation matrix through the resident equilibrium. Similarly, selection strength ($s$) and background recombination rate $(r)$ enter every realized matrix and therefore shape the entire random product. When realizations of $r_{12,t}$ occur on both sides of $r_{11}$ with positive probability, varying these parameters can shift the balance of contributions across generations and reverse the sign of the Lyapunov exponent. Parameters that have no effect on the direction of recombination rate evolution in the constant model thus become decisive under stochastic transmission.

The implications are immediate for how recombination-rate variability is treated empirically. Environmentally induced changes in recombination—with temperature, stress, season, or other conditions—have been documented across taxa and are often treated as noise around a fixed mean \cite{plough_effect_1917,stern_effect_1926,modliszewski_elevated_2018}. Our results show that, even if such fluctuations have no detectable effect on viability selection at the major loci, they can nonetheless alter evolutionary trajectories. What matters is not only whether the \emph{mean} recombination rate is lower than the resident rate, but how probability mass of this rate is distributed relative to the resident threshold and how extreme values are realized. Temporal structure in transmission is therefore itself a potential cause of evolutionary change in recombination.

More broadly, the analysis underscores a general point about selection on transmission modifiers. Evolutionary theory often attributes long-run change to viability selection acting on fixed transmission rules. Here transmission, affected by external factors, alters evolutionary outcomes without altering viability. Selection alone does not determine trajectory; it acts through, and is modulated by, the structure of transmission. When transmission varies, evolution depends on histories, not just on instantaneous selective comparisons. The Reduction Principle remains valid within its deterministic domain. Outside that domain, external modulation of recombination can qualitatively redirect evolutionary change.

Several extensions are natural and would sharpen the connection to data. Allowing temporal autocorrelation in recombination regimes would separate the effects of variance from those of persistence, which should matter whenever invasion dynamics are multiplicative. Relaxing the assumption of synchronous, population-wide fluctuations would permit analysis of partial synchrony or individual-level heterogeneity, bringing the model closer to empirical designs. Finally, extending the analysis beyond rare-modifier invasion to the joint dynamics of modifier frequency and linkage disequilibrium under stochastic transmission would address whether the regimes that favor invasion also sustain long-run polymorphism at the modifier locus. With $M_2$ randomly affecting the recombination rate, the dynamics of $M_2$ after invasion will be affected by $r_{22}$ produced by $M_2 M_2$, If $r_{22}$ is random, we might expect very complicated interactions between $r_{12, t}$ and $r_{22,t}$.

Selection on recombination rate modifiers cannot, in general, be inferred from constant-transmission arguments or from mean recombination rates alone. When transmission fluctuates temporally, selection acts on compounded histories, and parameters that are neutral in constant environments can become decisive. Secondary selection on transmission modifiers becomes a much more complex phenomenon.

\newpage
\section*{Acknowledgments}

We thank Shripad Tuljapurkar for numerous insightful and productive discussions. 

This research was supported in part by the Center for Computational, Evolutionary and Human Genomics (CEHG) at Stanford University.

\newpage
\appendix
\section{Frequencies of gametes produced by genotypes}
\label{appx:freqs_genotypes}
Table~\ref{tab:freqs} lists the frequencies of gamete types produced by each diploid genotype.
\begin{table}[htbp]
\centering
\caption{Frequencies of gametes produced by diploid genotypes. Each individual carries a pair of haplotypes; for a genotype formed by haplotypes $i$ and $j$, $w_{ij}$ denotes its fitness (with $w_{ij}=w_{ji}$). Entries give the probabilities of gamete types produced by each genotype.}
\label{tab:freqs}
\resizebox{\textwidth}{!}{
\renewcommand{\arraystretch}{1.5}
\LARGE
\begin{tabular}{|l|*{9}{c|}}
\hline
 \cline{2-9} & \textbf{$ABM_1$} & \textbf{$AbM_1$} & \textbf{$aBM_1$} & \textbf{$abM_1$} & \textbf{$ABM_2$} & \textbf{$AbM_2$} & \textbf{$aBM_2$} & \textbf{$abM_2$} \\ \hline
$ABM_1$ x $ABM_1$ & 1 & 0 & 0 & 0 & 0 & 0 & 0 & 0\\ \hline
$AbM_1$ x $ABM_1$ & $1/2$ & $1/2$ & 0 & 0 & 0 & 0 & 0 & 0\\ \hline
$aBM_1$ x $ABM_1$ & $1/2$ & 0 & $1/2$ & 0 & 0 & 0 & 0 & 0\\ \hline
$abM_1$ x $ABM_1$ & $\frac{1-r_{11}}{2}$ & $\frac{r_{11}}{2}$  & $\frac{r_{11}}{2}$ & $\frac{1-r_{11}}{2}$ & 0 & 0 & 0 & 0 \\ \hline
$AbM_1$ x $AbM_1$ & 0 & 1 & 0 & 0 & 0 & 0 & 0 & 0\\ \hline
$AbM_1$ x $aBM_1$ & $\frac{r_{11}}{2}$ & $\frac{1-r_{11}}{2}$ & $\frac{1-r_{11}}{2}$ & $\frac{r_{11}}{2}$ & 0 & 0 & 0 & 0\\ \hline
$AbM_1$ x $abM_1$ & 0 & $1/2$ & 0 & $1/2$ & 0 & 0 & 0 & 0\\ \hline
$aBM_1$ x $aBM_1$ & 0 & 0 & 1 & 0 & 0 & 0 & 0 & 0\\ \hline
$aBM_1$ x $abM_1$ & 0 & 0 & $1/2$ & $1/2$ & 0 & 0 & 0 & 0\\ \hline
$abM_1$ x $abM_1$ & 0 & 0 & 0 & 1 & 0 & 0 & 0 & 0\\ \hline \hline \hline

$ABM_2$ x $ABM_2$ & 0 & 0 & 0 & 0 & 1 & 0 & 0 & 0\\ \hline
$AbM_2$ x $ABM_2$ & 0 & 0 & 0 & 0 & $1/2$ & $1/2$ & 0 & 0\\ \hline
$aBM_2$ x $ABM_2$ & 0 & 0 & 0 & 0 & $1/2$ & 0 & $1/2$ & 0\\ \hline
$abM_2$ x $ABM_2$ & 0 & 0 & 0 & 0 & $\frac{1-r_{22}}{2}$ & $\frac{r_{22}}{2}$  & $\frac{r_{22}}{2}$ & $\frac{1-r_{22}}{2}$\\ \hline
$AbM_2$ x $AbM_2$ & 0 & 0 & 0 & 0 & 0 & 1 & 0 & 0\\ \hline
$AbM_2$ x $aBM_2$ & 0 & 0 & 0 & 0 & $\frac{r_{22}}{2}$ & $\frac{1-r_{22}}{2}$ & $\frac{1-r_{22}}{2}$ & $\frac{r_{22}}{2}$\\ \hline
$AbM_2$ x $abM_2$ & 0 & 0 & 0 & 0 & 0 & $1/2$ & 0 & $1/2$\\ \hline
$aBM_2$ x $aBM_2$ & 0 & 0 & 0 & 0 & 0 & 0 & 1 & 0\\ \hline
$aBM_2$ x $abM_2$ & 0 & 0 & 0 & 0 & 0 & 0 & $1/2$ & $1/2$\\ \hline
$abM_2$ x $abM_2$ & 0 & 0 & 0 & 0 & 0 & 0 & 0 & 1\\ \hline \hline \hline \hline

$ABM_1$ x $ABM_2$ & $1/2$ & 0 & 0 & 0 & $1/2$ & 0 & 0 & 0\\ \hline
$ABM_1$ x $AbM_2$ & $\frac{1-r}{2}$ & $\frac{r}{2}$ & 0 & 0 & $\frac{r}{2}$ & $\frac{1-r}{2}$ & 0 & 0\\ \hline
$ABM_1$ x $aBM_2$ & $\frac{r_{12} r + (1-r)(1-r_{12})}{2}$ & 0 & $\frac{r_{12}(1-r) + r(1-r_{12})}{2}$ & 0 & $\frac{r_{12}(1-r)+ r(1-r_{12})}{2}$ & 0 & $\frac{r_{12} r + (1-r)(1-r_{12})}{2}$ & 0\\ \hline
$ABM_1$ x $abM_2$ & $\frac{(1-r)(1-r_{12})}{2}$ & $\frac{r_{12}r}{2}$ & $\frac{(1-r)r_{12}}{2}$ & $\frac{r(1-r_{12})}{2}$ & $\frac{r(1-r_{12})}{2}$ & $\frac{(1-r)r_{12}}{2}$  & $\frac{r_{12}r}{2}$ & $\frac{(1-r)(1-r_{12})}{2}$\\ \hline
$AbM_1$ x $ABM_2$ & $\frac{r}{2}$ & $\frac{1-r}{2}$ & 0 & 0 & $\frac{1-r}{2}$ & $\frac{r}{2}$ & 0 & 0\\ \hline
$AbM_1$ x $AbM_2$ & 0 & $1/2$ & 0 & 0 & 0 & $1/2$ & 0 & 0\\ \hline
$AbM_1$ x $aBM_2$ & $\frac{r r_{12}}{2}$ & $\frac{(1-r)(1-r_{12})}{2}$ & $\frac{r(1-r_{12})}{2}$ & $\frac{r_{12}(1-r)}{2}$ & $\frac{r_{12}(1-r)}{2}$ & $\frac{r(1-r_{12})}{2}$ & $\frac{(1-r)(1-r_{12})}{2}$ & $\frac{r r_{12}}{2}$ \\ \hline
$AbM_1$ x $abM_2$ & 0 & $\frac{r r_{12}+(1-r)(1-r_{12})}{2}$ & 0 & $\frac{(1-r)r_{12}+r(1-r_{12})}{2}$ & 0 & $\frac{(1-r)r_{12}+r(1-r_{12})}{2}$ & 0 & $\frac{r r_{12}+(1-r)(1-r_{12})}{2}$\\ \hline
$aBM_1$ x $ABM_2$ & $\frac{(1-r)r_{12}+r(1-r_{12})}{2}$ & 0 & $\frac{r r_{12} + (1-r)(1-r_{12})}{2}$ & 0 & $\frac{r r_{12}+(1-r)(1-r_{12})}{2}$ & 0 & $\frac{(1-r)r_{12}+r(1-r_{12})}{2}$ & 0\\ \hline
$aBM_1$ x $AbM_2$ & $\frac{(1-r)r_{12}}{2}$ & $\frac{r(1-r_{12})}{2}$ & $\frac{(1-r)(1-r_{12})}{2}$ & $\frac{r r_{12}}{2}$ & $\frac{r r_{12}}{2}$ & $\frac{(1-r)(1-r_{12})}{2}$ & $\frac{r(1-r_{12})}{2}$ & $\frac{r_{12}(1-r)}{2}$\\ \hline
$aBM_1$ x $aBM_2$ & 0 & 0 & $1/2$ & 0 & 0 & 0 & $1/2$ & 0\\ \hline
$aBM_1$ x $abM_2$ & 0 & 0 & $\frac{1-r}{2}$ & $\frac{r}{2}$ & 0 & 0 & $\frac{r}{2}$ & $\frac{1-r}{2}$\\ \hline
$abM_1$ x $ABM_2$ & $\frac{r(1-r_{12})}{2}$ & $\frac{(1-r)r_{12}}{2}$ & $\frac{r r_{12}}{2}$ & $\frac{(1-r)(1-r_{12})}{2}$ & $\frac{(1-r)(1-r_{12})}{2}$ & $\frac{r r_{12}}{2}$ & $\frac{r_{12}(1-r)}{2}$ & $\frac{r(1-r_{12})}{2}$\\ \hline
$abM_1$ x $AbM_2$ & 0 & $\frac{(1-r)r_{12}+r(1-r_{12})}{2}$ & 0 & $\frac{r r_{12}+(1-r)(1-r_{12})}{2}$ & 0 & $\frac{r r_{12}+(1-r)(1-r_{12})}{2}$ & 0 & $\frac{(1-r)r_{12}+r(1-r_{12})}{2}$\\ \hline
$abM_1$ x $aBM_2$ & 0 & 0 & $\frac{r}{2}$ & $\frac{1-r}{2}$ & 0 & 0 & $\frac{1-r}{2}$ & $\frac{r}{2}$\\ \hline
$abM_1$ x $abM_2$ & 0 & 0 & 0 & $1/2$ & 0 & 0 & 0 & $1/2$\\ \hline
\end{tabular}}
\end{table}
\section{Centrosymmetric structure and block diagonalization of the invasion Jacobian} \label{app:blockdiag}
This appendix derives equations \eqref{eq:JplusJminus} from the explicit centrosymmetric form of the
invasion Jacobian \eqref{eq:J-centro}. The objective is to show how the $4$-dimensional linear invasion system separates into two independent $2$-dimensional subsystems.

The Jacobian obtained in Section~\ref{sec:constant-recomb} is
\[
J=
\begin{pmatrix}
\mathcal{A} & \mathcal{B} & \mathcal{C} & \mathcal{D}\\
\mathcal{E} & \mathcal{F} & \mathcal{G} & \mathcal{H}\\
\mathcal{H} & \mathcal{G} & \mathcal{F} & \mathcal{E}\\
\mathcal{D} & \mathcal{C} & \mathcal{B} & \mathcal{A}
\end{pmatrix}.
\tag{\ref{eq:J-centro}}
\]
Rows $1$ and $4$ are mirror images of each other, as are rows $2$ and $3$, and the same symmetry holds for columns. Equivalently, reversing the haplotype order $(5,6,7,8)\mapsto(8,7,6,5)$ leaves $J$ unchanged. This entrywise invariance is precisely the definition of \textit{centrosymmetry}.

This structure implies that sums and differences of reversal-paired haplotypes evolve independently. To verify this, add the first and fourth equations:
\begin{align*}
\hat{\bar w}(x_5'+x_8')
&=(\mathcal A x_5+\mathcal B x_6+\mathcal C x_7+\mathcal D x_8)
 +(\mathcal D x_5+\mathcal C x_6+\mathcal B x_7+\mathcal A x_8)\\
&=(\mathcal A+\mathcal D)(x_5+x_8)+(\mathcal B+\mathcal C)(x_6+x_7).
\end{align*}
The last equality is obtained by grouping coefficients: in the sum, $x_5$ and $x_8$ both have total coefficient $\mathcal A+\mathcal D$, and $x_6$ and $x_7$ both have total coefficient $\mathcal B+\mathcal C$. Because members of each reversal paired pair enter with equal total weight, no term proportional to the pairwise difference (e.g.\ $x_5-x_8$) can appear in the sum equation.

Subtracting the same two equations yields
\begin{align*}
\hat{\bar w}(x_5'-x_8')
&=(\mathcal A x_5+\mathcal B x_6+\mathcal C x_7+\mathcal D x_8)
 -(\mathcal D x_5+\mathcal C x_6+\mathcal B x_7+\mathcal A x_8)\\
&=(\mathcal A-\mathcal D)(x_5-x_8)+(\mathcal B-\mathcal C)(x_6-x_7),
\end{align*}
in which the sums $(x_5+x_8)$ and $(x_6+x_7)$ drop out by the same coefficient-matching argument. Performing the analogous addition and subtraction for the second and third rows of \eqref{eq:J-centro} gives the corresponding relations for $x_6'\pm x_7'$. Hence, the dynamics of the two sums are closed, and the dynamics of the two differences are closed.

Motivated by these cancellations, we make the linear change of variables:
\[
y_1=x_5+x_8,\quad y_2=x_6+x_7,\quad
y_3=x_5-x_8,\quad y_4=x_6-x_7,
\qquad
\mathbf y=(y_1,y_2,y_3,y_4)^\top.
\]
In matrix form, $\mathbf y=T\mathbf x_{\mathrm{inv}}$ with
\[
T=
\begin{pmatrix}
1 & 0 & 0 & 1\\
0 & 1 & 1 & 0\\
1 & 0 & 0 & -1\\
0 & 1 & -1 & 0
\end{pmatrix}.
\]
Solving for $\mathbf x_{\mathrm{inv}}$ gives $x_5=\tfrac12(y_1+y_3)$, $x_8=\tfrac12(y_1-y_3)$, $x_6=\tfrac12(y_2+y_4)$, $x_7=\tfrac12(y_2-y_4)$, and therefore
\[
T^{-1}=\frac12
\begin{pmatrix}
1 & 0 & 1 & 0\\
0 & 1 & 0 & 1\\
0 & 1 & 0 & -1\\
1 & 0 & -1 & 0
\end{pmatrix},
\]
consistent with \eqref{eq:Tandinv}.
Substituting $\mathbf x_{\mathrm{inv}}=T^{-1}\mathbf y$ and multiplying by $T$ on the left yields $\hat{\bar w}\,\mathbf y'=(TJT^{-1})\,\mathbf y$. The closure relations above imply that $TJT^{-1}$ contains no coupling between $(y_1,y_2)$ and $(y_3,y_4)$, so the transformed system is block diagonal:
\[
\hat{\bar w}\,\mathbf y'=
\begin{pmatrix}
J^+ & 0\\
0   & J^-
\end{pmatrix}
\mathbf y,
\]
where the blocks are read off from \eqref{eq:J-centro} as
\[
J^+=
\begin{pmatrix}
\mathcal A+\mathcal D & \mathcal B+\mathcal C\\
\mathcal E+\mathcal H & \mathcal F+\mathcal G
\end{pmatrix},
\qquad
J^-=
\begin{pmatrix}
\mathcal A-\mathcal D & \mathcal B-\mathcal C\\
\mathcal E-\mathcal H & \mathcal F-\mathcal G
\end{pmatrix}.
\]
The block $J^+$ governs the evolution of the total invader frequency within each reversal-paired haplotype class, while $J^-$ governs the evolution of differences within each pair.

If the modifier allele is introduced without a directional bias between reversal-paired haplotypes—that is, without preferentially placing $M_2$ on one major-locus phase rather than its phase-reversed counterpart—then the initial frequencies satisfy
\[
x_5(0)=x_8(0), 
\qquad 
x_6(0)=x_7(0).
\]
Here $(5,8)$ and $(6,7)$ denote the two pairs of phase-reversed (coupling/repulsion) $M_2$-bearing haplotypes, which are dynamically equivalent under the transmission–selection recursion. In the transformed coordinates $\mathbf y = T \mathbf x_{\mathrm{inv}}$, the components $y_3$ and $y_4$ represent the antisymmetric contrasts within each reversal pair (proportional to $x_5-x_8$ and $x_6-x_7$, respectively). Thus the above equalities are equivalent to
\[
y_3(0)=y_4(0)=0.
\]

Because the linearized recursion is invariant under independent exchange of the members of each reversal pair, the symmetric and antisymmetric subspaces are invariant under the Jacobian. Consequently,
\[
(y_3(t),y_4(t)) \equiv (0,0)
\quad \text{for all } t,
\]
so the dynamics remain confined to the symmetric subspace. Invasion is therefore determined by the dominant eigenvalue of the restriction of the Jacobian to this subspace, namely $J^+/\hat{\bar w}$, as in Section~\ref{sec:constant-recomb}.

\newpage
\section*{Code and supplementary material}

All code and supplementary material used for the analyses in this paper are available at
\url{https://github.com/elisaheinrichmora/Stochastic_Transmission_Recombination.git}.

\bibliography{references} 
\bibliographystyle{abbrv}
\end{document}